\documentclass[fleqn,usenatbib]{mnras}

\usepackage{newtxtext,newtxmath}

\usepackage[T1]{fontenc}

\DeclareRobustCommand{\VAN}[3]{#2}
\let\VANthebibliography\thebibliography
\def\thebibliography{\DeclareRobustCommand{\VAN}[3]{##3}\VANthebibliography}

\usepackage{graphicx}	
\usepackage{gensymb}
\usepackage{amsmath}	
\usepackage{amssymb}	
\usepackage{indentfirst}

\title[The structure of the co-orbital stable regions]{The structure of the co-orbital stable regions as a function of the mass ratio}

\author[L. Liberato \& O. Winter]{
L. Liberato$^{1}$\thanks{E-mail: luana.l.mendes@unesp.br} and
O. C. Winter$^{1}$\thanks{E-mail: othon.winter@unesp.br}
\\
$^{1}$ UNESP - Sao Paulo State University, Grupo de Dinamica Orbital e Planetologia, CEP 12516-410, Guaratingueta, SP, Brazil}

\date{Accepted XXX. Received YYY; in original form ZZZ}

\pubyear{2020}

\begin{document}
\label{firstpage}
\pagerange{\pageref{firstpage}--\pageref{lastpage}}
\maketitle

\begin{abstract}
Although the search for extra-solar co-orbital bodies has not had success so far, it is believed that they must be as common as they are in the Solar System. Co-orbital systems have been widely studied, and there are several works on stability and even on formation. However, for the size and location of the stable regions, authors usually describe their results but do not provide a way to find them without numerical simulations, and, in most cases, the mass ratio value range is small. In the current work, we study the structure of co-orbital stable regions for a wide range of mass ratio systems and built empirical equations to describe them. It allows estimating the size and location of co-orbital stable regions from a few system’s parameters. Thousands of massless particles were distributed in the co-orbital region of a massive secondary body and numerically simulated for a wide range of mass ratios ($\mu$) adopting the planar circular restricted three-body problem. The results show that the horseshoe regions upper limit is between $9.539\times10^{-4}<\mu<1.192\times10^{-3}$, which correspond to a minimum angular distance from the secondary to the separatrix between $27.239\degree$ and $27.802\degree$. We also found that the limit to exist stability in the co-orbital region is about $\mu=2.3313\times10^{-2}$, much smaller than the value predicted by the linear theory. Polynomial functions to describe the stable region parameters were found, and they represent estimates of the angular and radial widths of the co-orbital stable regions for any system with $9.547\times10^{-5}\leq\mu\leq2.331\times10^{-2}$.

\end{abstract}

\begin{keywords} methods: numerical - celestial mechanics - minor planets, asteroids, general - planets and satellites: dynamical evolution and stability.

\end{keywords}


\setlength{\parskip}{0.1em}

\section{Introduction}
\begin{figure}
    \centering
    \includegraphics[scale=0.82]{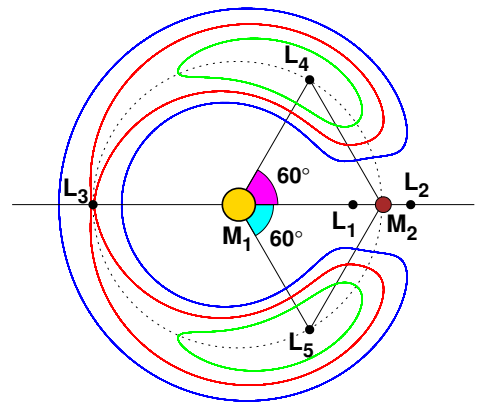}
    \caption{Location of the five Lagrangian points, represented by the black dots, and orbits around them in the rotating frame. The blue path is horseshoe orbit, and the green paths are tadpole orbits. The red line represents the separatrix between the two regimes. The yellow circle represents the central body $ M_ {1} $ and the brown circle $ M_ {2} $.}
    \label{fig:lpoints}
\end{figure}

Leonhard Euler (1707-1783) and Joseph-Louis Lagrange (1736-1813) studied the restricted three-body problem. They found five equilibrium points that hold their positions in the frame rotating with the secondary body, the well known Lagrangian equilibrium points (Figure \ref{fig:lpoints}).

The $L_1$, $L_2$ and $L_3$ points were discovered first by Euler and published in 1788 entitled \textit{Moturium Corporum se Mutuo Attrahentium Super Eadem Linea Recta}. They are linearly unstable and are known as co-linear points because they lay on the imaginary line that connects the primary body to the secondary one. The last two points, $L_4$ and $L_5$, were discovered and published between 1867 and 1892 by \citeauthor{lagrange}. They are known as triangular points because they are on the vertices of the equilateral triangles formed with the primary and secondary bodies. \cite{gascheau1843mouvements} proved that the triangular points would only be linearly stable if $\mu<0.0385$, where $\mu=M_2/(M_1+M_2)$ is the mass ratio of the system. 

Through the results found by Lagrange, in 1911 Ernest Brown studied the system Sun-Jupiter-trojans and proved that bodies could have stable and periodic orbits around the $L_3$, $L_4$ and $L_5$ \citep{brown1911}, the so-called tadpole and horseshoe orbits.

Tadpole is the orbit around one of the triangular points $L_4$ or $L_5$ and has this name because, in the rotating frame, it has a shape that looks like a tadpole (Figure \ref{fig:lpoints}). For the horseshoe orbits, however, these are paths around $L_3$, $L_4$ and $L_5$. Between both types of orbits, there is a separatrix that, as the name suggests, separates the two orbit regimes. Mathematically it is an imaginary line, but in reality, it has a width and turns the region near the separatrix chaotic.

It was only in 1906 that Max Wolf discovered the first co-orbital asteroid confirming Lagrange’s work. It was called 588 Achilles and was found librating around $L_4$ point of Jupiter’s orbit. After Achilles, thousands of other small bodies were found sharing Jupiter's orbit.

Today it is known that, besides Jupiter, there are also co-orbital bodies to Earth, Venus, Mars and Neptune. Moreover, there are several systems of co-orbital satellites around Saturn such as Janus-Epimetheus, Telesto-Calypso-Thetis and Polydeuces-Helene-Dione. As these bodies were being discovered, many important works were made to study the co-orbital systems. There are studies on the dynamics in a restricted three-body problem \citep{danby1964stability,dermott1981a,lohinger1993stability,mittal2020analysis}, on the stable regions around Lagrangian points \citep{deprit1967stability,marchal1991predictability,erdi2005stability,erdi2007secondary,zhou-li-dvorak2009,cuk2012long}, studies on the co-orbital regions of Solar System's planets \citep{mikkola1992numerical,nesvorny2002long,scholl2005instability,tabachnik2000asteroids}, dynamics of some known co-orbital satellites \citep{dermott1981b,yoder1983,treffenstadt2015}, on stability and formation of the Solar System's trojans \citep{izidoro2010,Donnison1985,chanut2008nebular,mourao2006,pitjeva2019masses,zhou2011dynamics,zhou2019orbital,zhou2020systematic,dvorak-lhota-2012,mikkola1990studies,lykawka2009origin,lykawka2011origin,marzari2013long}, and even hypothetical co-orbital exoplanets \citep{schwarz2009dynamics,schwarz2012stability,schwarz2015possibility,beauge2007,dvorak2012trojans}.

Since the discovery of the first exoplanet around a main-sequence star, made by \citet{mayor&queloz1995}, the technology has allowed a huge increase in the number of known exoplanets in the last few years, and now there are more than four thousand confirmed exoplanets \footnote{exoplanet.eu}. It is already known that non-planetary structures were found in extra-solar systems, like comets \citep{Kiefer2014}, an asteroid belt \citep{moro2008}, rings around exoplanets \citep{exo-ring2015} and more recently a moon in a formation process \citep{exomoon2019}. So, although the search projects have not found co-orbital bodies until now \citep[e.g.][]{troyproject}, it is believed that they may be as common as they are in our system and they will probably be detected soon. 

Hence, we made a numerical study on the co-orbital stability region for a range of mass ratio compatible with exoplanets values. The systems are composed of two massive bodies in planar circular orbits and massless particles in the co-orbital region. The goal of this study is to measure some parameters to determine the boundaries of stable regions accurately. For that, in section \ref{sec:simul} is described in detail the studied systems and the numerical simulations that were performed. In section \ref{sec:results} are shown the results and measured parameters. And finally, in section \ref{sec:conc} are presented our final remarks and comments about the present work. 

\begin{figure*}
  \centering
  \includegraphics[width=\linewidth,height=8.5cm]{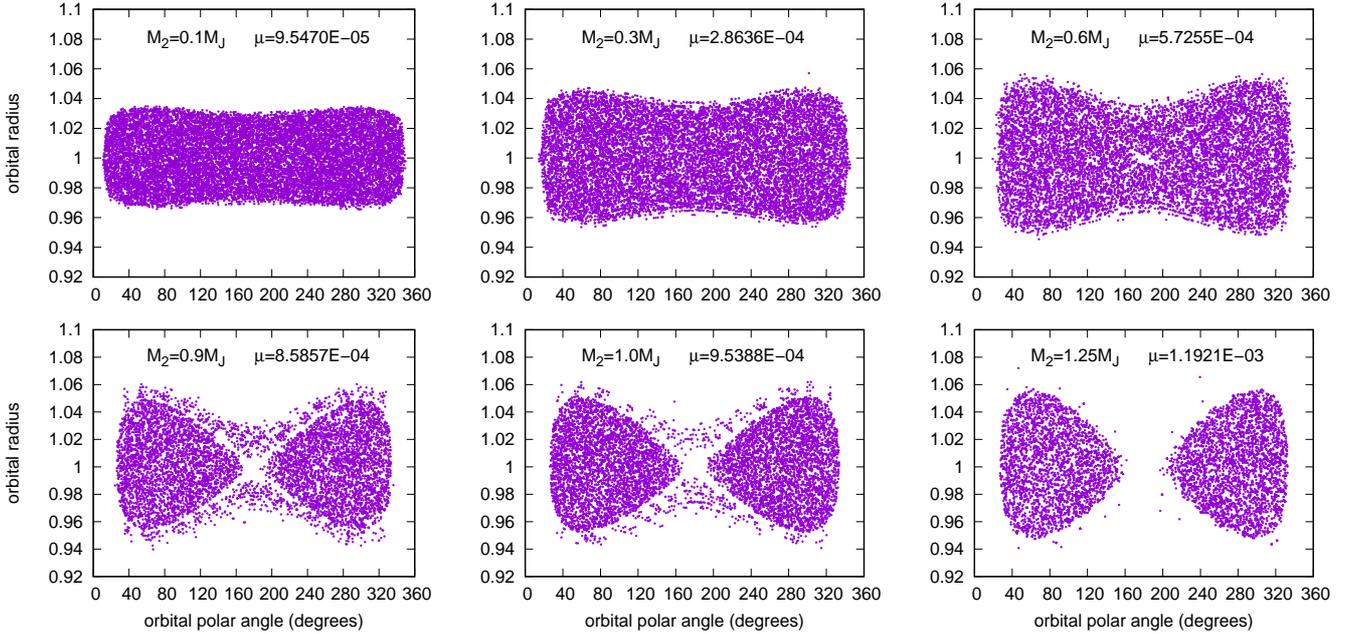}
  \caption{Evolution of the stable co-orbital region for small mass ratio values. The plots show the polar angle versus the orbital radius for six different values of mass ratio of the system. The dots represent the initial conditions of the particles that survived until the end of the simulations.}
  \label{ferradura}
\end{figure*}
\section{Simulations}
\label{sec:simul}

In the studied systems, there is one massive central body named $M_1$, one massive secondary body, $M_2$, orbiting the primary in a circular orbit, and a coplanar ring of thousands of particles distributed in the co-orbital region. The numerical simulations were performed with Mercury, a package of integrators for the N-body problem \citep{chambers1999}, using the Bulirsh-Stoer integrator. 

Following the steps of \citep{dermott1981a}, we normalize to one the sum of the mass ratio from both massive bodies and $M_2$ semi-major axis, $a$. The massless particles were distributed with a semi-major axis from $a - W_{chaos}$ to $a+ W_{chaos}$ using an uniformly randomized distribution, where $W_{chaos}=2.1$ $a$ $\mu^{(1/3)}$ measures the width of the chaotic region due to the overlap of first-order resonances in the vicinity of $M_2$ \citep{wisdom1980resonance}. The orbits of all bodies are initially planar and circular, and the central body was assumed to be a Sun-like star with the Sun's mass, which can be scaled for other systems using the mass ratio.

The simulations were integrated up to $7\times10^5$ orbital periods of $M_2$ and were divided into two groups. In the first group, the mass of $M_2$ varied from $0.1$ to $1.0$ Jupiter's masses, with a step of $0.1 M_J$. For each value of mass, there are $50,000$ particles in the co-orbital region angularly distributed from $0\degree$ to $360\degree$. In the second group, there are $25,000$ particles angularly distributed between $0\degree$ to $180\degree$ in the co-orbital region of each system. $M_2$ assumed values from $0.1$ to $1.0$ Jupiter's masses, with a step of $0.1 M_J$ (as in the first group) and from $1.25M_J$ to $24M_J$, with a step of $0.25M_J$. 

From preliminary studies, we noticed that this two groups separation was more convenient because it divides the orbit regimes and favours to study the tadpole separately from the horseshoe orbits. As it is theoretically known, the co-orbital region is symmetric around the $M_1M_2$ axis, so the results obtained for the region around $L_4$ will be equal to those around $L_5$. Therefore, to save a large amount of CPU time, we focus only on the region around $L_4$. Hence, if the angular difference between a particle and $M_2$ is larger than $180\degree$, then it is considered to be in a chaotic trajectory and is removed from the simulation. This assumption is because particles in tadpole orbits will remain around the same Lagrangian point through the entire simulation. With this consideration, only the stable particles that are in the tadpole region, around $L_4$ will survive. For all simulations, in both groups, it was assumed that the particles that survive for $7\times10^5$ orbital periods of the secondary are in stable orbits. 

It was implemented, in the Mercury package, two conditions of particles removal from the system: if a particle is less than $8\degree$ away from $M_2$, or if it crosses the internal or external radial limits set on $a\pm 1.1W_{chaos}$. Once again, the value $8\degree$ was obtained from preliminary simulations. From those, it was noticeable that all particles within this condition were ejected from the co-orbital region or collided with $M_2$. Also, these delimitations decrease the CPU time required for each simulation, and they eliminate particles in chaotic trajectories from the co-orbital region. Altogether 615 numerical simulations were performed.
\section{Results and Discussions}
\label{sec:results}
Since the systems are planar and during the simulation the particles acquire eccentricities smaller than $10^{-3}$, it is assumed that all orbits are nearly circular and so the results are expressed in polar coordinates, where $r$ is the orbital radius and $\Theta$ the polar angle. It is considered that $M_2$ is at $0\degree$, so the angle $\Theta$ of the particles also means the angular separation from $M_2$.

We are interested in the stable co-orbital regions. It means to know the boundaries of the places near the orbit of $M_2$ where a particle will remain until the end of the numerical simulation. The final location of the particle is not relevant for this work. Based on simulation conditions that were mentioned previously, it can be ensured that if the particle survived then it is inside the co-orbital region. So the results are based on the initial conditions of all particles that reached the end of the simulation.

The first thing noticed from the results is that the horseshoe orbits are only accessible for the cases with mass ratio values $\mu\leq9.539\times10^{-4}$, as illustrated in Figure \ref{ferradura}. For larger mass ratios, all the particles in horseshoe orbits were ejected. As expected from the theory \citep{dermott1981a}, the smaller the mass ratio, the greater its ability to hold particles in horseshoe orbits for a long period. Therefore, there is a mass ratio limit for particles to be in horseshoe orbits and this limit is about $1 M_J$, or $\mu\approx10^{-3}$. 
 \setlength{\parskip}{0.1em}

\begin{figure*}
  \centering
  \includegraphics[width=\textwidth,height=8cm]{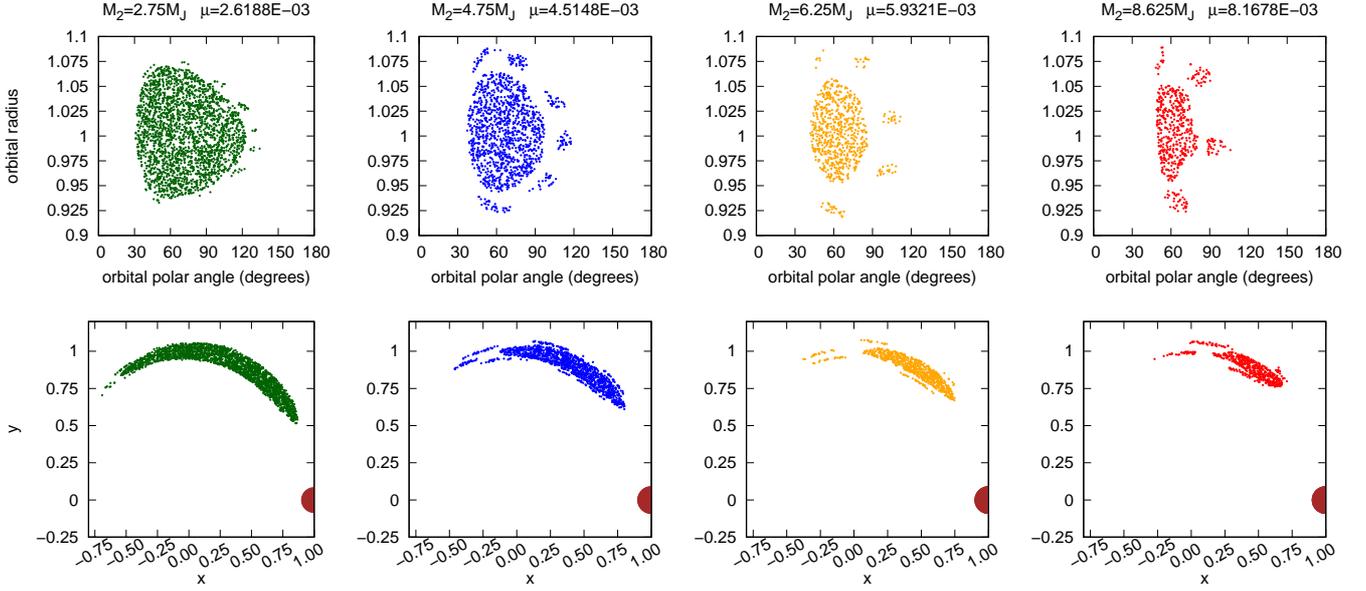}
  \caption{Regions of stability in the co-orbital region. The top plots show the polar angle versus the orbital radius of the surviving particles for four different values of mass ratio. The bottom plots show the same surviving particles in the x-y plane. The dots represent the initial conditions of the particles that survived until the end of the simulations, and the circle represents the location of $M_2$.}
  \label{tail}
\end{figure*}
In Figure \ref{tail} is noticeable that as $\mu$ increases, gaps are opened inside the stable regions creating some islands where particles can survive until the end of the simulation. Looking at the actual shape of these regions, it can be noticed that these structures look like tails of the main stable region. The same structures were found in other numerical studies \citep[e.g.][]{lohinger1993stability,sandor2003symplectic,schwarz2007survey,erdi2005stability}. Those are islands of stability around the main stable region, and they are associated with resonances of librational motions \citep{erdi2007secondary}.

The changes in sizes of the stable regions imply that the number of surviving particles at the end of the simulations follows a similar behaviour. The larger the stable region, more particles can survive, while the smaller the stable region, fewer particles survive. But this growth and decrease of the stable region's size is not a linear function of the mass ratio, as we will show in the following subsections, where are presented the results and discussion for the cases with $\mu\leq9.539\times10^{-4}$ and the cases with $\mu>9.539\times10^{-4}$.
\subsection{Horseshoe Stable Regions}
\label{sec:ang-sep}

\begin{figure}
 \includegraphics[scale=1.2]{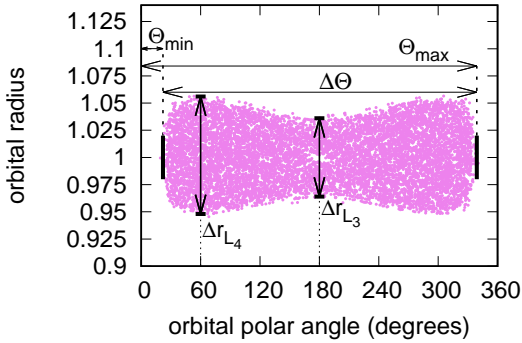} \caption{Example of how the radial ($\Delta r_{L_3}$ and $\Delta r_{L_4}$) and angular ($\Theta_{min}$, $\Theta_{max}$ and $\Delta\Theta$) measurements were made for the cases with horseshoe stable regions. That is a plot of orbital polar angle versus radius of the surviving particles in the co-orbital region. The dots represent the initial conditions of the particles that survived until the end of the simulation.}

  \label{fig:med-horseshoe}
\end{figure}

\begin{figure}
        \centering
          \includegraphics[scale=1]{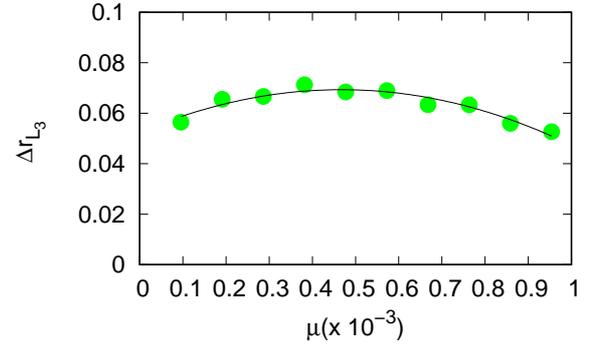}
          \caption{The radial width of the stable region near $L_3$, for horseshoe orbits, as a function of the mass ratio. The circles indicate the measured values of the radial width for each case of mass ratio and the solid line indicates the quadratic fit given by Equation \eqref{eq-fit} and Table \ref{tab:coef_horse}.}
            \label{horseshoe_l3}
        \end{figure}

\begin{figure}
        \centering
        \includegraphics[scale=1]{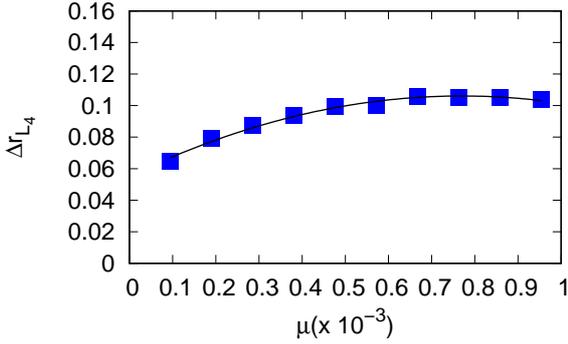}
        \caption{The radial width of the horseshoe stable region near $L_4$ as a function of the mass ratio. The squares represent the measured values of the radial width for each case of mass ratio, and the solid line shows the quadratic fit given by Equation \eqref{eq-fit} and Table \ref{tab:coef_horse}}
        \label{horseshoe_l4}
\end{figure}

\begin{figure}
        \centering
        \includegraphics[scale=1]{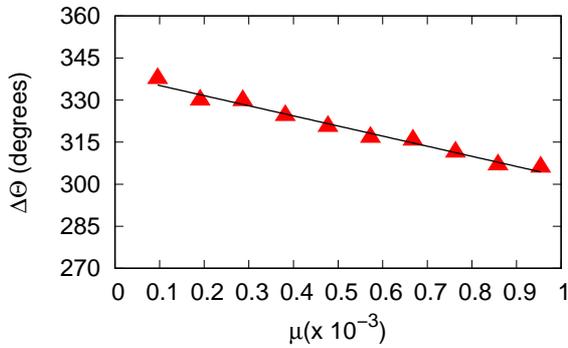}
        \caption{Angular width of the horseshoe stable region as a function of the mass ratio. The triangles indicate the measured value of the angular width for each case of mass ratio. The solid line shows the linear fit given by Equation \eqref{eq-fit} and Table \ref{tab:coef_horse}.}
        \label{horseshoe_ang}
\end{figure}

As discussed previously, it is possible to separate the mass ratio values where particles in horseshoe orbits can and cannot be found. So in this section, only the results for the first group of simulations called horseshoe stable regions, with $\mu\leq9.539\times10^{-4}$ ($M_2\leq 1 M_J$), are analysed. Three parameters are measured: radial width near $L_4$ ($\Delta r_{_{L_4}}$), radial width near $L_3$ ($\Delta r_{_{L_3}}$) and angular width near the semi-major axis of $M_2$ ($\Delta\Theta=\Theta_{\rm max}- \Theta_{\rm min}$), as shown in Figure \ref{fig:med-horseshoe}. The results of these measurements are summarized in Figures \ref{horseshoe_l3}, \ref{horseshoe_l4} and \ref{horseshoe_ang}.

To measure the radial width near $L_3$ ($\Delta r_{L_3}$), first, all particles that have their initial $\Theta$ value between $177\degree$ and $183\degree$ are selected, so it can be assured that they are in the vicinity of $L_3$, located at $180\degree$. Then, with the selected particles, the third most external and third most internal values of orbital radius are chosen, this way it is taken into account only the stable particles (avoiding any eventually spurious single particle). So, these values are considered to be the radial boundaries of the stable region near the $L_3$ point. In Figure \ref{horseshoe_l3} are shown the measured values of $\Delta r_{L_3}$ as a function of the mass ratio. The solid line presents the quadratic fit done according to Equation \eqref{eq-fit} and Table \ref{tab:coef_horse}:
\begin{equation}
        f(\mu)=\sum_{i=0}^{N} c_{i}\mu^{i},
        \label{eq-fit}
\end{equation}
        
where $N$ is the order of the polynomial fit, $c_i$ are the fitting coefficients. This equation was used to do the fitting of all parameters studied in the present work. Therefore, $f(\mu)$ provides the boundary of each studied parameter which, in the horseshoe stable regions, can be the radial width near $L_4$, radial width near $L_3$ or the angular width $\Delta \Theta$, depending on the coefficients used.

The function coefficients and reduced $\chi^2$ value \footnote{$\chi^2$ is a parameter used to measure the discrepancy between observed values and the values expected under the model in question. The reduced form is the $\chi^2$ value divided by the degrees of freedom of the model. The closest this value is to $1$, the better is the fit.} for each fit are presented in Table \ref{tab:coef_horse}.The fitting shows that the evolution of $\Delta r_{L_3}$ has a parabolic behaviour, despite the small variation. It means that the stable regions near $L_3$ grow radially as the $\mu$ value increases, until about $\mu\approx0.5\times10^{-3}$, where it has its peak and then goes decreasing until disappear. 
 For the radial width near $L_4$, all the particles with a polar angle between $57\degree$ and $63\degree$ were separated and, just like the previous measurement, it is taken the third most external and third most internal values of orbital radius, considering these as the radial boundaries of the stable region near $L_4$. Figure \ref{horseshoe_l4} shows that as the mass ratio is increased, the larger the radial width near the point $L_4$ become. \citet{murraylivro} showed that for very small mass ratio values (about $10^{-8}$) the horseshoe orbits have an estimated radial half-width of $r_{H}\propto a\mu^{1/3}$, recalling that $a$ is the semi-major axis of $M_2$. We also found an empirical equation to describe this parameter. Once again the results were fitted to a polynomial function (Equation \eqref{eq-fit}), shown as the line in Figure \ref{horseshoe_l4}, and the coefficients are given in Table \ref{tab:coef_horse}. In this measurement, there is also a parabolic behaviour, but as expected, the stable regions around $L_4$ does not disappear when $\mu$ value gets closer to $10^{-3}$.         

The other parameter measured for the horseshoe regions is the angular width of the stable region($\Delta \Theta$). For that, it was applied the same approach used in the previous measurements. All particles with an orbital radius between 0.98 and 1.02 were selected, and it was taken the third larger ($\Theta_{max}$) and third smaller ($\Theta_{min}$) values of the polar angles, considering these as the angular limits of the horseshoe stable region. The results presented in Figure \ref{horseshoe_ang} show that the angular width of the horseshoe stable region tends to shrink as the mass ratio increases, and this behaviour is as smooth as the radial width near $L_4$, as expected. In this case, a linear fitting was enough to describe the behaviour of the results. The coefficients and reduced $\chi^2$ values are shown in Table \ref{tab:coef_horse}.
 
For the small values of $\mu$ that were studied, the co-orbital stable regions presented slight changes in their size and shape, allowing simple functions to be fitted and describe their properties.

\begin{table}
  \centering
  \begin{tabular}{|p{1.4cm}|p{1.7cm}|p{1.7cm}|p{2cm}|}
    \multicolumn{4}{c}{Parameters of horseshoe Regions}\\\hline\hline
    &    $f(\mu)=\Delta r_{_{L_4}}$    &    $f(\mu)=\Delta r_{_{L_3}}$    &    $f(\mu)=\Delta \Theta$  [$\degree$]    \\ \hline
    c0  &   5.552$\times10^{-2}$ & 5.245$\times10^{-2}$ & 338.808 \\ \hline
    c1    &    132.229 & 72.246 &-3.608$\times10^{4}$ \\ \hline
    c2    &    -8.588$\times10^{4}$ &-7.737$\times10^{4}$ &----------- \\ \hline
    \hline
    Reduced $\chi^2$&    1.428 & 1.428 & 1.250 \\
    \hline
  \end{tabular}
  \caption{Table of coefficients obtained by the polynomial adjust of the parameters studied from horseshoe stable regions, and reduced $\chi^2$ value for each parameter fitted.}
  \label{tab:coef_horse}
\end{table}
\subsection{Tadpole Stable Regions}
\begin{figure}
    \centering
    \includegraphics[scale=1.2]{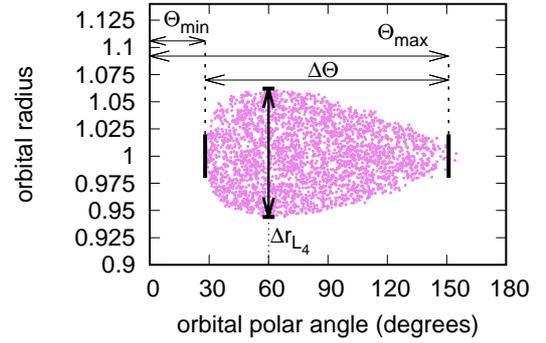}
    \caption{Example of how the radial ($\Delta r_{L_4}$) and angular ($\Theta_{min}$, $\Theta_{max}$ and $\Delta\Theta$) measurements were made for the cases with tadpole stable regions. Plot of orbital polar angle versus orbital radius of the surviving particles in the co-orbital region near $L_4$. The dots represent the initial conditions of the particles that survived until the end of the simulation.}
    \label{fig:med-tadpole}
\end{figure}

In this section are presented the results of tadpole stable regions. Using the same methods as before, we also measured the radial width near $L_4$, and the minimum and maximum $\Theta$ (Figure \ref{fig:med-tadpole}). We highlight that this study was made focusing only on $L_4$ because there is symmetry between the triangular Lagrangian points. Hence, the results found for one of them must be similar to the other. 
 
\begin{figure}
        \centering
        \includegraphics[scale=1]{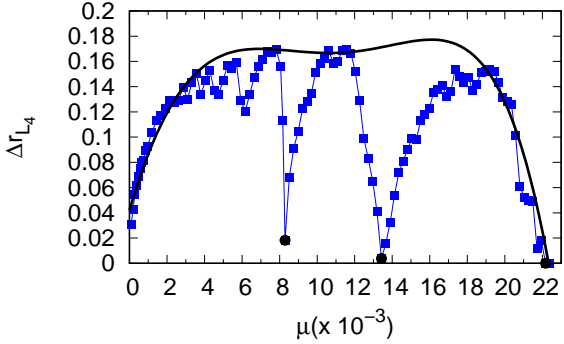}
        \caption{The radial width of the stable region near $L_4$ as a function of the mass ratio. The squares indicate the measured value of the radial width for each simulated case of $\mu$. The black circles represent the three critical $\mu$ values and the polynomial fit is presented as the black solid line (see more details in the text).}
        \label{ajuste_r}
\end{figure}

\begin{figure}
        \centering
        \includegraphics[scale=1]{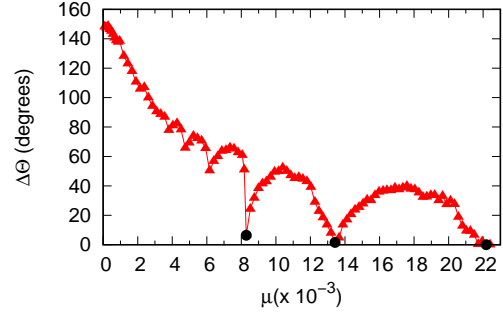}
        \caption{The angular width of the stable region near $L_4$ as a function of the mass ratio. The triangles indicate the measured value of the angular width for each simulated case of $\mu$ value. The black circles represent the three critical $\mu$ values.}
        \label{angular_width}
\end{figure}

Figure \ref{ajuste_r} shows that, at first, the radial width tends to grow as $\mu$ values increase, but for values larger than $\approx3\times10^{-3}$ it stops growing and begins to oscillate. Figure \ref{angular_width} shows that the angular width has the opposite tendency, where it decreases before presenting the same wavering behaviour as the radial width. These variations mean that the co-orbital stable region is expanding and shrinking as the value of $\mu$ is increased.        

 In the work of \citet{danby1964stability} the author analytically studied the stability of the triangular Lagrangian points. To do so, he starts with the premise that a particle in a stable motion around one of the triangular points will not suffer energy loss. Therefore, he studies the behaviour of a particle to which is given a small displacement from its initial position around a Lagrangian point. After some manipulation, the linearized equations of motion are presented in a matrix form. 
 
 To obtain the solution of the equations of motion it is needed to diagonalize the transformation matrix. From that are obtained four eigenvectors and four eigenvalues. These eigenvalues ($\lambda$), which depend only on $\mu$, can be interpreted as the particles frequencies of motion in its oscillations around $L_4$( or $L_5$). The condition for a motion to be stable, in the cases of small eccentricity, is that all four $\lambda$ values are unequal and purely imaginary. The instability in a particle's motion arises from the fact that certain values of $\mu$ add a real part to some of the $\lambda$ values. The author found that there are a few values of $\mu$ that lead particles to unstable motions, but he lets clear that the values were a raw approximation. 
 
 Further, \cite{deprit1967stability} showed that, for small values of eccentricity, there are four values of $\mu$ that turn the co-orbital region around the Lagrangian points into unstable. These values are $\mu_1=3.8521\times10^{-2}$  (same as Gascheau limit), $\mu_2=2.4293\times10^{-2}$,$\mu_3=1.3516\times10^{-2}$ and $\mu_4=1.0913\times10^{-2}$. The authors mathematically found these values through a series of calculations. Meanwhile, our results were found through numerical simulations. The critical mass ratio values that we find are $\mu_2=2.2173\times10^{-2}$, $\mu_3=1.3423\times10^{-2}$ and $\mu_4=0.8285\times10^{-2}$, indicated by the black circles in Figures \ref{ajuste_r} and \ref{angular_width}. The differences between the critical $\mu$ values from \cite{deprit1967stability} are probably due to analytical approximations and the fact that we have data for specific values of $\mu$.
 
In Figures \ref{ajuste_r} and \ref{angular_width} is possible to see that, besides these obvious minima, there are other noticeable values of $\mu$ for which the tadpole regions shrink. That behaviour is associated with resonances of librational motions. The particles moving around $L_4$ have two frequencies of motion: one for short and one for long period librations. But the Lagrangian point also has frequencies of motion around the system's barycentre. So, the combination of all those frequencies causes resonances that affect the co-orbital region \citep{erdi2007secondary}, which explains the oscillatory behaviour in the results.
 
\begin{figure}
        \includegraphics[scale=1.1]{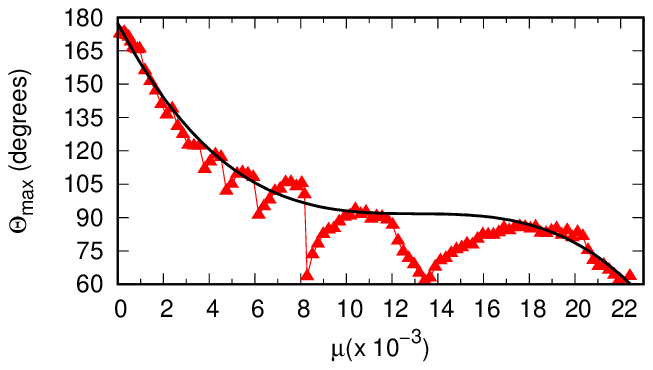}
        \includegraphics[scale=1.1]{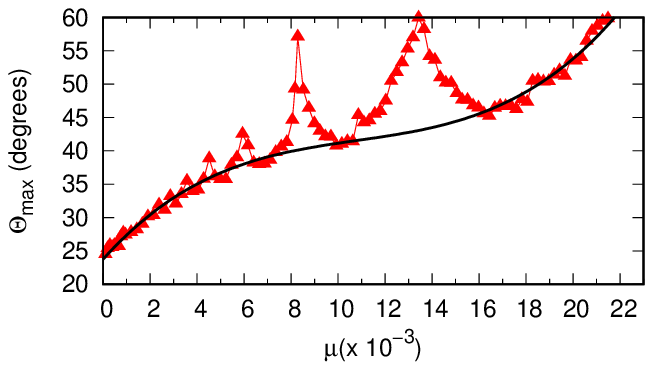}
        \caption{Maximum (top) and minimum polar angle (bottom) of the stable regions as a function of the mass ratio. The triangles represent the measured values of the polar angle for each case of $\mu$. The polynomial fits are presented as the black solid lines (see more details in the text).}
        \label{ajuste_theta}
\end{figure}

  Disregarding the valleys and peaks, i.e. considering a system without the resonances mentioned previously, the polynomial function in Equation \eqref{eq-fit} was fitted to our data. In these cases $f(\mu)$ can be the maximum radial width near $L_4$ (Figure \ref{ajuste_r}), the minimum polar angle $\Theta_{min}$ (Figure \ref{ajuste_theta} bottom) or the maximum polar angle $\Theta_{max}$ (Figure \ref{ajuste_theta} top). The values of these coefficients for each fit are indicated in Table \ref{tab:coef}. 
 
 In Figure \ref{ajuste_r}, the polynomial fit is presented as the black line. From that fit, the maximum radial size of a tadpole stable region can be reasonably estimated, even though it does not precisely describe the measured    values. 
 In Figure \ref{ajuste_theta} are shown two plots of mass ratio versus maximum (top) and minimum polar angle (bottom), and the black lines in each one show the fitting curve made with the polynomial function (Equation \eqref{eq-fit}). Again it can be obtained approximated values for the minimum and maximum polar angles and then estimate the maximum angular width of the tadpole regions.
\subsection{Mass Ratio Limits}
 \cite{yoder1983} studied the dynamics of the co-orbital Saturn-Janus-Epimetheus system. In their work, the authors do an analytic approximation to the motion of the satellites with accuracy to mass ratio of order $10^{-8}$. After some calculation, they present an equation (Equation (10) in \cite{yoder1983}) that defines the equivalent energy constant. With a few manipulations, we find Equation \eqref{eq-yoder}. This equation is used to obtain the angular distance $\Theta$ between two bodies ($M_2$ and $M_3$) with comparable masses sharing the same orbit, but very small masses when compared to the central body ($M_1$):
\begin{equation}
    -\frac{sin\left(\frac{\Theta}{2}\right)}{3\epsilon^2 n^2}\left(\frac{d\Theta}{dt}\right)^2=4sin^{3}\left(\frac{\Theta}{2}\right) + \frac{2E}{\epsilon^2 n^2}sin\left(\frac{\Theta}{2}\right)+1,
    \label{eq-yoder}
\end{equation} 
where $E$ is the equivalent constant energy, $\Theta$ is the angular distance between $M_2$ and $M_3$, $\epsilon^2 = (M_3 + M_2) / M_1$, and $n$ is the average mean motion. 

\begin{figure}
        \centering
        \includegraphics[scale=1.5]{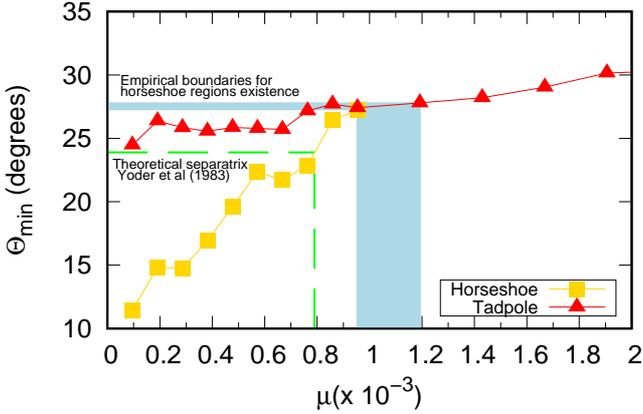}
        \caption{Minimum polar angle as a function of the mass ratio. The dashed line represents the theoretical separatrix $\Theta_{min}$ value \citep{yoder1983} and the blue area represents the range of $\Theta_{min}$ values where the empirical boundaries for horseshoe regions existence is within. Triangles and squares represent $\Theta_{min}$ for tadpole and horseshoe regions, respectively.}
        \label{separatrix}
\end{figure}

Considering the separatrix between tadpole and horseshoe regimes, at the maximum distance between $M_2$ and $M_3$, $\Theta=180\degree$  and $\frac{d\Theta}{dt}=0$, then the left side of Equation \eqref{eq-yoder} is null and it is found that the energy at the separatrix is given by $E=-\frac{5\epsilon^2 n^2}{2}$. However, at the minimum separation points $\frac{d\Theta}{dt}=0$ is also valid, and using the separatrix energy in Equation \eqref{eq-yoder}, we obtain:
\begin{equation}
    4sin^{3}\left(\frac{\Theta_{min}}{2}\right) + 5sin\left(\frac{\Theta_{min}}{2}\right)+1=0
    \label{eq-min}
\end{equation}
From Equation \eqref{eq-min} it is found that the separatrix of a system with this configuration has its minimum separation at $23.9\degree$. It means that, for systems with a mass ratio of order $10^{-8}$, tadpole orbits can not be found with $\Theta_{min}<23.9\degree$, only horseshoe orbits. That agrees with the results presented in Figure \ref{separatrix}, where we highlight that all tadpole regions have their minimum angular distance larger than the separatrix location predicted by \cite{yoder1983}.

Is also noticeable that for some $\mu$ values, horseshoe regions have their minimum angular distance beyond $23.9 \degree$, proving that the angular location of the separatrix depends on the mass ratio of the system.

In addition, Figure \ref{separatrix} shows that the mass ratio boundary for the existence of horseshoe stable regions is between $9.539\times10^{-4}$ ($M_2=1 M_J$) and $1.192\times10^{-3}$ ($M_2=1.25$ $M_J$), which leads to an angular distance upper limit between $27.239\degree$ and $27.802\degree$. The angular distance lower limit found for horseshoe regions is around $11\degree$. However, this might be due to the range of mass ratio values used in this study, and this value may not represent the boundary.

 We have also performed simulations for a secondary body's mass between $25M_J$ and $30M_J$. For mass ratio values between $2.217\times10^{-2}\leq\mu\leq2.785\times10^{-2}$ the co-orbital region appears highly affected by resonances of librational motions. Hence, that might be the mass ratio limit to exist stability in the co-orbital region. However, previous works have shown that it is possible to exist stability in the co-orbital region of systems with much greater mass ratio values than those we have studied, such as trojan planets in binary star systems \citep[e.g.][]{schwarz2015possibility}.
\vspace{-0.3cm}\section{Final Remarks}
\label{sec:conc}
 Our main goal in the current work was to study co-orbital stable regions, to analyse their characteristics and measure their boundaries for different values of the mass ratio ($\mu$) of each studied system. For $\mu\leq9.539\times10^{-4}$, which was called horseshoe stable regions, it was seen that the width near $L_3$ does not change significantly as $\mu$ increases, but it has a parabolic behaviour with a peak near $\mu = 5 \times 10 ^{-4}$. On the contrary, it was noticed that as $\mu$ is increased, $\Delta\Theta$ tends of decreasing while $\Delta r_{_{L_4}}$ increases. 

 For the tadpole stable regions, it was found an oscillating behaviour of the parameters. These oscillations are associated with resonances of librational motions, which significantly reduce the size of the stability regions, and in some cases even makes them disappear. We identified these critical mass ratio values and compared with those obtained from previous works. We noticed that as smaller critical mass ratio more the theoretical and the empirical values coincide.

 A fit was made with a polynomial function considering the data in a scenario neglecting the oscillations and, even though the curves do not describe in details the results, they give an approximation of the upper limit values for the measured parameters. Overall, were found functions to describe the behaviour of co-orbital stability and to estimate, approximately, their size and location depending only on the mass ratio of a co-orbital system. 

 From the results, we found that the upper limit for the horseshoe orbits to be accessible for particles in the system is slightly greater than the expected from the theory \citep{yoder1983}. We found a range of $\Theta_{min}$ values where the separatrix must be within and the corresponding range of $\mu$ values. 

Lastly, our simulations for $2.217\times10^{-2}\leq\mu\leq2.785\times10^{-2}$ showed that the entire co-orbital region is unstable because no particles survived. Thus, the limit for linear stability predicted by previous works are much greater than what our results showed, and the instability in the co-orbital region might begin for much smaller mass ratio values than the expected.
\begin{table}
 \centering
 \begin{tabular}{|p{0.4cm}|p{2cm}|p{2.2cm}|p{2.2cm}|}
   \multicolumn{4}{c}{Parameters of tadpole Regions}\\\hline\hline
   &    $f(\mu)=\Delta r_{_{L_4}}$    &    $f(\mu)=\Theta_{min}$ [$\degree$]    &    $f(\mu)=\Theta_{max}$ [$\degree$]\\ \hline
   c0  &   4.29204$\times10^{-2}$ & 23.7555 & 177.446 \\\hline
   c1    & 5.10548$\times10^{1}$&3.9306$\times10^{3}$ & -1.9521$\times10^{4}$ \\ \hline
   c2    &    -6.94081$\times10^{3}$ &-3.1692$\times10^{5}$ & 1.4872$\times10^{6}$ \\ \hline
   c3    &    3.31355$\times10^{5}$ &9.7771$\times10^{6}$ & -3.7911$\times10^{7}$ \\ \hline
   c4    &    0.00000 &------------- &----------- \\ \hline
   c5   &    -2.56434$\times10^{8}$ &------------ &----------- \\ \hline
   \hline
 \end{tabular}
 \caption{Table of coefficients obtained by the polynomial adjust of the parameters studied from tadpole stable regions.}
 \label{tab:coef}
\end{table}
\vspace{-0.3em}
\section*{Acknowledgements}
This research was financed in part by the Coordenação de
Aperfeiçoamento de Pessoal de Nível Superior - Brasil (CAPES) - Finance Code 001, Fundação de Amparo à Pesquisa do Estado de
São Paulo (FAPESP) - Proc. 2016/24561-0, Conselho Nacional
de Desenvolvimento Científico e Tecnológico (CNPQ) - Proc. 305210/2018-1.
\vspace{-0.4em}
\section*{Data availability}
The data underlying this article will be shared on reasonable request to the corresponding author.



\bibliographystyle{mnras}
\bibliography{bibliografia} 





\bsp	
\label{lastpage}
\end{document}